\documentclass{aa}  

\usepackage{graphicx}
\usepackage{txfonts}
\usepackage{xcolor}
\begin{document} 

\title{The network analysis of the cosmic web as a tool to constrain cosmology and cosmic magnetism }



\author{A. Rudakovskyi\inst{1,3}, F. Vazza\inst{1}, M. Tsizh\inst{1,4}}

\offprints{%
 E-mail:}

\institute{Dipartimento di Fisica e Astronomia, Universit\'{a} di Bologna, Via Gobetti 93/2, 40122, Bologna, Italy
\and INAF-Istituto di Radioastronomia, Via Gobetti 101, 40129, Bologna, Italy
\and Bogolyubov Institute for Theoretical Physics of the NAS of Ukraine, Metrolohichna Str. 14-b, Kyiv, 03143, Ukraine
\and Astronomical Observatory of Ivan Franko National University of Lviv, Kyryla i Methodia str. 8, Lviv, 79005, Ukraine }
\authorrunning{A Rudakovskyi et al.}
\titlerunning{Primordial magnetism}
\date{Accepted ???. Received ???; in original form ???}

 
  \abstract
   {The spatial distribution of haloes in the Cosmic Web encodes a wealth of information about the underlying cosmological model. These haloes can be represented as nodes of a graph, whose structural properties reflect cosmological parameters.
      
   }
   {Using our new \texttt{MAKITRA} suite of cosmological magneto-hydrodynamical simulations covering a total volume of $(300 \rm ~Mpc)^3$ and with $21$ physical model variations (including variations of $\sigma_8$ and of different models of primordial magnetic fields, PMFs), we investigate the sensitivity of network-based statistics describing the Cosmic Web to variations in cosmological and PMF scenarios.}
{We focus on several complementary metrics that characterise the spatial distribution of dark and baryonic matter haloes: two-point correlation functions, network-centrality statistics, and counts-in-cell measurements. We first compare the halo–halo correlation functions across different cosmological models. For the network analysis, we represent haloes as vertices of the Cosmic Web and compute multiple centrality measures, whose cumulative distributions we evaluate for universes with varying PMF strengths. Finally, we quantify halo abundances within randomly placed spheres of fixed radius to assess differences between scenarios.} 
   {First, we find that the statistics of the centralities of the network can serve as a novel sensitive probe of the cosmological parameter $\sigma_8$. Moreover, we find that this network analysis approach can allow us to distinguish the presence of PMFs with initial strength $\simeq 4\,\mathrm{nG}$ from the scenarios with much weaker PMFs. }
   {}
   \maketitle
%

\section{Introduction}

 Investigating the distribution of galaxies in the Cosmic Web has long been one of the fundamental and earliest methods for constraining cosmological models.
Indeed, galaxies are a key tracer of the cosmic matter distribution across redshift, and in combination with advanced numerical simulations, it is in principle possible to account for their bias and use them to infer cosmological parameters in a robust way. 
This can be addressed with various approaches. The most commonly used approach to characterise the statistical properties of galaxies (and of their dark matter halos) in the cosmic web is the two-point correlation function (2ptCF) \citep[see, e.g.][]{Peebles:1980, Landy:1993}. This statistic has been widely used to constrain cosmological parameters, such as the matter density parameter $\Omega_m$ and the initial amplitude of fluctuations of matter density $\sigma_8$ \citep[see, e.g.,][]{Cole2005, MacCrann2018}. However, the distribution of galaxies in the cosmic web cannot be fully captured by the 2ptCF alone. In fact, points generated through Levy flights can mimic the galaxy-galaxy correlation function from \texttt{IllustrisTNG} simulations, while exhibiting a quite different distribution \citep{Hong:2016}.

To improve parameter constraints beyond those obtainable from the galaxy 2ptCF alone, several alternative approaches have been proposed. For example: multipole expansions of the 2ptCF in redshift space \citep[see, e.g.][]{Kaiser:1987}, marked power spectra \citep[see, e.g.][]{Cowell2024, Marinucci24}, higher-order correlation functions \citep[e.g., a bispectrum,][]{Croton:2004,Sefusatti:06}, counts-in-cell (CIC) statistics \citep[][]{White:79, Uhlemann:20}, among others. We use the latter in this paper. CIC statistics have previously been used to estimate the bias parameter between dark and baryonic matter \citep{Salvador2018, Repp:2020}, to constrain the neutrino mass \citep{Uhlemann:20}, and to investigate clustering processes \citep{Sanjuan2015}.

Alternatively, a natural way to describe the distribution of galaxies in the cosmic web is through various higher-order statistical and topological analyses of the point clouds formed by their spatial positions. One can consider the population of galaxies as tracing the nodes of a complex network. At the most fundamental level, the Cosmic Web may be described as a network, where nodes (representing the galaxies) are connected by edges if the distance between them is less than a certain linking length. There are at least two possible ways to take advantage of such an approach: firstly, one can characterise the evolution of the Cosmic Web in a holistic way, by comparing its network metrics for different cosmologies, cosmic times, realisations, etc. For example, \citet{Hong:2020} have demonstrated that transitivity and average neighbourhood degree seem to be promising diagnostics for discriminating different fractions of matter $\Omega_m$ and the dark energy equation of state parameters $w$. Before this, \citet{Ueda2003} applied a graph-theoretical approach to distinguish between different cosmologies in simulations. On the other hand, one can look for the correlations between network metrics and physical and/or observable quantities that characterise the galaxies. The examples of the latter type of studies are \citet{Hong:2015} and \cite{deRegt:2017}, where the authors applied network analysis to the \texttt{COSMOS} galaxy catalogue and demonstrated correlations between network centralities and the colours or evolutionary parameters of galaxies. Another relevant example is \citet{Tsizh:2020}, where we investigated the relationships between the node network metrics and the types of structures (voids, filaments, sheets, and superclusters) to which the nodes belong. Network analysis can help discriminate between observational and mock catalogues \citep{Hong2018}. The alternative way, persistent homology, offers a scale-free approach to pointcloud and network analysis, achieved by considering all linking lengths through a sequence of simplex complexes. The halo distribution in simulations also including primordial magnetic fields was already analysed in \citet{Tsizh23} with such methodology, showing that it can discriminate between $\sigma_8$ variations. Finally, \citet{Ouellette2025} combines both of the approaches mentioned here (CIC / neighbour count statistic and persistent homology) to address the problem of $\sigma_8$-$\Omega_m$ degeneracy in cosmological observation.

Network analysis also has the potential of identifying additional non-gravitational effects on the evolution of the cosmic web. 
In this work, we are mainly focused on the possible effect of Primordial Magnetic Fields (PMFs), whose existence is getting increasingly more consistent with a number of observations.   
The detection of magnetic fields diffuse in the extreme periphery of clusters of galaxies \citep[e.g.][]{2019Sci...364..981G,2022SciA....8.7623B,2022Natur.609..911C,2025A&A...696A.203P} and in cosmic filaments \citep[e.g.][]{os20, vern21, vern23, Carretti24} implies the presence of magnetisation processes in addition to what the can be released by galaxies. Moreover, the non-detection of inverse Compton emission from blazars at $\sim 1-10 \rm GeV$ energies has been to infer a lower limit ($\geq 10^{-17}-10^{-15} \rm G$ ) on the magnetization of voids in the local, $z \leq 0.2$, Universe \citep[e.g.][]{2010Sci...328...73N,2021Univ....7..223A}, again incompatible with the sole magnetisation from galactic processes \citep[e.g.][]{2021MNRAS.505.5038A,2024ApJ...963..135T,2025arXiv251026918G}. 
Finally, the state-of-the-art analysis of the Cosmic Microwave Background (CMB) also provides stringent upper limits from the possible shape of the power-spectrum of PMFs, although presently both "causal" (i.e. with magnetic energy peaking at small scales) or ``inflationary" (i.e. with magnetic energy peaking at scales larger than the cosmic horizon) are allowed \citep[][]{2019JCAP...11..028P}, although the most recent analysis of radio data from different telescopes combined with theoretical models suggests that PMFs of the inflationary kind are slightly favoured \citep[][]{Carretti24,2024ApJ...977..128M,cava25}.
The theoretical expectations on the generation and properties of PMFs are still very uncertain \citep[e.g.][]{2013A&ARv..21...62D,sub16,2020arXiv201010525V}, and the potential impact of PMFs on cosmology is manifold. If present already at recombination (or before), PMFs can induce matter inhomogeneities on top of the standard cosmological structure formation scenario, accelerating cosmic recombination \citep[e.g.][]{2019PhRvL.123b1301J} and speed up the formation of the first dwarf galaxies \cite[see, e.g.][]{Kim:94, Kahniashvili:2012dy, AdiTal:2023, Ralegankar:2024}, which in turn can lead to earlier reionisation by UV photons from their stars. The number counts of dwarf galaxies \citep{Sanati:20, Sanati:24}, Lyman-$\alpha$ forest \citep{Shaw:2012, Kahniashvili:2012dy, Pavicevic:2025}, line intensity mapping constraints \citep{Adi:2023}, and upcoming observations of 21-cm emission from neutral hydrogen \citep[see, e.g.,][]{Cruz:24, Bhaumik:25} can be used to constrain or detect the presence of PMFs at the dawn of galaxy formation. On the other hand, after recombination, PMFs can cause additional pressure which effectively increases the Jeans length and suppresses the formation of halos below a certain mass scale \citep[see, e.g. ][sec II.D  and references therein]{Cruz:24}. 

All this considered, simulating in detail how structure formation and PMFs co-evolve is an important, yet under-looked, line of research which is worth exploring in detail, given the steadily increasing coverage and sensitivity of large galaxy surveys already operating in
in the optical and near infra-red spectrum \citep[e.g.][]{Euclid:2024yrr}, or in the radio band \citep[e.g.][]{2022A&A...659A...1S,2024PASA...41...26D}, culminating with the 
upcoming Square-Kilometre Array \citep[e.g. SKA][]{2020PASA...37....2W}. 

In this work, we present a novel suite of magneto-hydrodynamical simulations, \texttt{MAKITRA}, designed to investigate the effects of varying cosmological parameters and PMFs on the structure of the cosmic web.  
This paper is organized as follows. Section \ref{sec:methods} presents our new suite of cosmological simulations, along with the various algorithms employed to identify matter halos, reconstruct their network, and measure their spatial properties. Sect. \ref{sec:results} presents our results on the relation between PMFs and network or clustering properties of simulated universes, while Sect.~\ref{sec:discussion} discusses limitations of our approach. Our conclusions are given in Sect.~\ref{sec:conclusions}, while the Appendix presents important supporting material for our suite of simulations, and additional network statistics not presented in the main paper. 

\section{Methods}
\label{sec:methods}

\begin{figure*}
\centering
\includegraphics[width=0.99\textwidth]{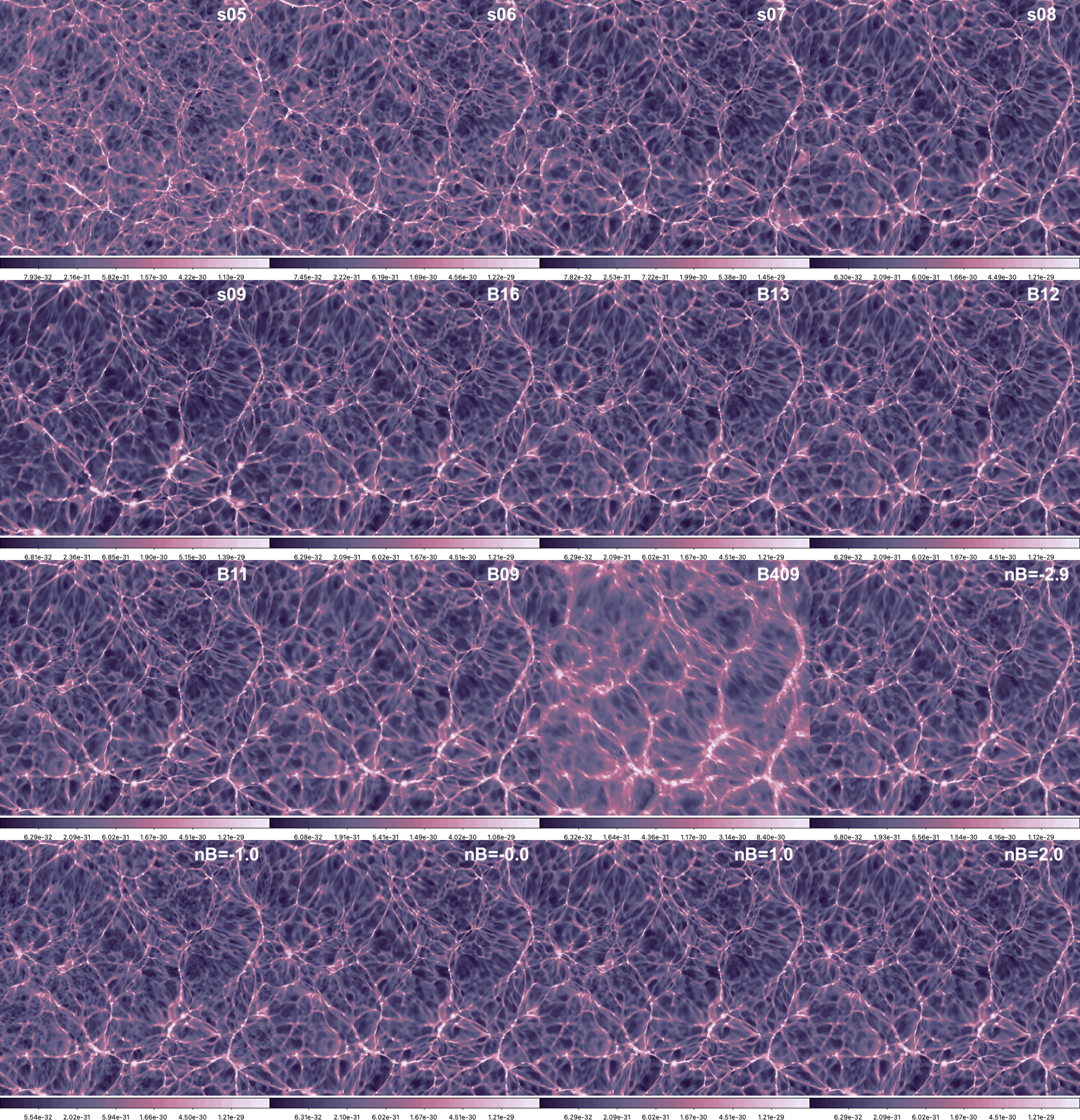}
\caption{Distribution of baryonic matter density for one of our simulated volumes, for 16 model variations (out of a total of 21 variations) explored in the \texttt{MAKITRA} suite at $z=0.2$. The side of each box is $150$ Mpc, and the colorbar gives the number density in [$\rm g/cm^3$].}
\label{fig:all_models}
\end{figure*}

\subsection{The \texttt{MAKITRA} suite of ENZO cosmological simulations}

We have produced a suite of ideal MHD simulations created using the Eulerian \texttt{ENZO} magneto-hydrodynamical code in a fixed-grid modality \cite[see, e.g.][]{Enzo:2014}. 

We simulated 8 independent cubic cosmic volumes ($150^3 \rm Mpc^3$ each) to sample a total volume of $300^3 \rm Mpc^3$, investigating for each of them $21$ different variations of $\sigma_8$, or in the amplitude and morphology of PMFs, or additionally for the presence of feedback effects from star formation and active galactic nuclei (even if the analysis of the latter is deferred to future work). In particular, in this paper, we are interested in constraining the impact of all model variations on the cosmic network, and on the growth of cosmic structures in general, and assessing to what extent such signatures can be observed with the statistics of real galaxy surveys of the sky.  Each box has a box length~$150 ~\rm Mpc$ (comoving), and a uniform cell resolution~$\Delta x=292.98\, \rm kpc$ (comoving). The Dark Matter is sampled with fixed mass resolution  $ 6.7\cdot10^9 M_{\odot}$ (even if this changes with cosmological variations of $\Omega_M$).
  Our baseline cosmological parameters are $\Omega_M=0.308$, $\Omega_\Lambda=0.692$, $\Omega_b=0.0478$, $\sigma_8=0.8$ and $H_0=67.8 \rm ~km/s/Mpc$. The initial redshift of all runs is $z_{in}=30$.  
 For all simulations, we included radiative gas cooling from a constant cosmic composition, and the effect of a uniform UV re-heating background on the temperature of baryons, introduced at $z=7$ and with a $L_\nu \propto \nu^{-1.57}$ UV emission spectrum from quasars, following \citet{HM12}. 
 
 The suite is called \texttt{MAKITRA}, \footnote{\url{https://cosmosimfrazza.eu/makitra}. A "Makitra" is a big bowl used to grind and mix ingredients, traditionally used in Ukrainian and Polish kitchens.} and it contains the following  physical variations: 

\begin{itemize}
\item in the ``Baseline model" (first entry in Table~\ref{tab_cosmo}) we used  $\sigma_8=0.8$, no feedback, and a primordial uniform magnetic field with amplitude $B_0=0.1 \rm ~nG$ (comoving). Therefore, what  we consider our baseline model is not the most unmagnetised model of the suite, but rather our fiducial one, based on recent complementary studies on the likely amplitude of magnetic fields in the Universe, based on radio observations \citep[e.g.][and references therein]{cava25};
\item variations in magnetic field amplitude for uniform magnetic field models $B_0 = 4 \cdot 10^{-9}, 2 \cdot 10^{-9}, 10^{-9}\,, 10^{-11}\,, 10^{-12}\,,10^{-13}\,,10^{-16}\,$G (B409, B209, B09, B11, B12, B13, B16 models hereafter). The initial seed of magnetic fields $B_0=10^{-16}\,$G produces present-day magnetic fields in voids which are on the lower end of what is allowed by high-energy observation of blazars \citep[e.g.][for a review]{2021Univ....7..223A}, while $B_0 \geq 2 \cdot 10^{-9}$ G models are on the high end of what is allowed for uniform fields by CMB analysis \citep[][]{1997PhRvL..78.3610B}.
\item variations of the initial topology of PMFs, for models with vectors randomly drawn from an initial power-law spectrum, $P_B(k)=P_0 k^{n}$ (see Sect.~\ref {A1}): $n=-2.9,-1.0,0.0,1.0,2.0$. The spectral normalisation $P_0$ scales with the standard deviation of the magnetic field within a given fixed length ($1 ~\rm Mpc$ comoving), and is based on the available constraints the analysis of magnetic effects on the Cosmic Microwave Background \citep[e.g.][]{PLANCK2015,2019JCAP...11..028P}, see Sect.~\ref{A2} for more details; 
\item variations of the initial amplitude of matter density fluctuations, referred to as usual to a smoothing scale of $8 \rm~ Mpc$ comoving:  $\sigma_8=0.5\,,0.6\,,0.7\,,0.8\,,0.9$ (s05, s06, s07, s08 and s09 models).
\item variations in which the additional release of thermal and magnetic field energy (in the initial form of dipoles) is injected by feedback events from active galactic nuclei, in addition to primordial magnetic fields of different amplitudes, to assess the role of "astrophysical" seeding scenarios of extragalactic magnetic fields. We mention this set of model variations for completeness in describing the \texttt{MAKITRA}  suite, even if the latter is not analysed in this work, but only used in the Appendix for preliminary testing. 

\end{itemize}

We've run each model variation on 8 different boxes, whose initial conditions were produced with ENZO starting from different random seed fields to assign the initial phases of the matter density and velocity perturbations, with the goal of assessing the role of cosmic variance on our results. This yielded a total of $168$ runs, and for each of them we saved 8 snapshots from $z=6$ to $z=0$ and the corresponding halo catalogues (see Sect.~\ref{halo_cat}) to compute network properties. 
In this work, we focus on the $z=0.2$ snapshot of all \texttt{MAKITRA} runs and model variations, on the basis that each of our $150^3 \rm Mpc^3$ comoving volume at that redshift covers about $12^\circ \times 12^\circ$ in the sky, which gives a reasonably large projected area to compare with what can be done by real optical/IR surveys of the sky. 

We provide full description of our simulations in Table~\ref{tab_cosmo} with the main parameter variations explored in \texttt{MAKITRA}  (also including variations not discussed in this paper), as well as a more in-depth overview of the choices adopted to model primordial magnetic fields in our models in the Appendix~\ref{A1}. 

Fig.~\ref{fig:all_models} gives the visual impression of the effect of the explored model variations on the distribution of baryonic matter in a thin slice through the middle plane of the same simulated $150^3 \rm Mpc^3$ box at $z=0.2$. The complementary distribution of magnetic fields explored in \texttt{MAKITRA}  is given in the Appendix~\ref{A1} (Fig.~\ref{fig:new_sim}).  
Fig.~\ref{fig:PDF_B} plots the distribution of the median magnetic field strengths for our runs (limited to those with variations of the initial magnetic field) at the same epoch and as a function of the baryon over-density. In this Figure we additionally show the existing approximate range of lower limits for the amplitude of PMFs, implied by the non-detection of inverse Compton cascade emission in blazars via $\gamma-$rays  \citep[e.g.][and references therein]{2021Univ....7..223A}, and the maximum magnetic field amplitude in filaments suggested by recent detection in the radio band, via synchrotron and Faraday Rotation \citep[e.g.][and references therein]{cava25}. Finally, we also mask out the high-density regime where our simulations cannot correctly capture
turbulent dynamo amplification, due to their limited resolution, which artificially keeps the numerical Reynolds number low within halos \citep[e.g.][and references therein]{review_dynamo}.  This shows that the variety of PMFs explored in this work spans a still large range of uncertainty for the magnetisation of large scale structures, based on observational constraints. This further motivates our investigation of the possible dynamical effect of PMFs on the cosmic network of halos, in the following Sections, as an additional way to constrain PMFs.

An initial version of the \texttt{MAKITRA} suite was produced on the JUWELS cluster in JUELICH  using \texttt{ENZO} v2.6, running on 64 nodes, using a total of ~200,000 core hours (BREAKTHRU Computing project with P.I. F. Vazza). The updated suite was produced on the LEONARDO cluster at CINECA, using the GPU-ported version of \texttt{ENZO} using about 100,000 core hours on 64 nodes (IscrB\_CREW project with P.I. F. Vazza).

\begin{figure}
    \includegraphics[width=0.495\textwidth]{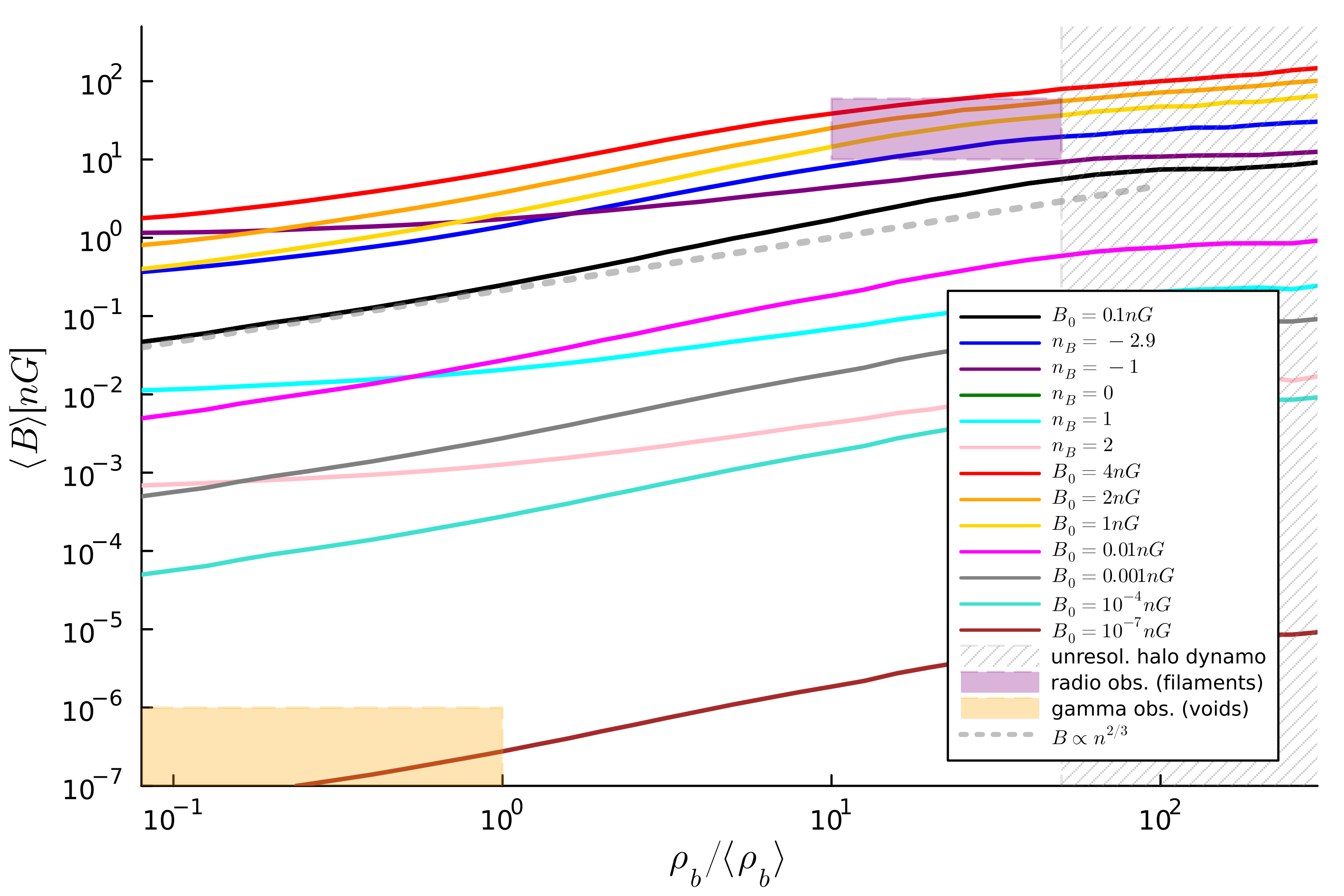}
    \caption{Distribution of the mean magnetic field strength as a function of the baryonic matter over density at $z=0.2$, for runs with magnetic field variations. The coloured areas show the approximate range of lower limits on PMFs from $\gamma$-ray observations, and the maximum magnetic field in filaments derived by radio observations, while the hatched area shows the approximate density range of clusters, where our simulation struggles to resolve dynamo amplification properly. The additional dashed line shows the $B\propto n^{2/3}$ scaling expected for pure isotropic compression of magnetic field lines.}
    \label{fig:PDF_B}
\end{figure}

\subsection{Halo finding and halo catalogues}
\label{halo_cat}
For each model and redshift in the suite, we extracted catalogues of halos based on a spherical overdensity criterion. In detail, we identified all high density peaks in the simulation (considering the total, $\rho_{DM}+\rho_B$, matter density) and then iterated the estimation of the average matter density enclosed within spherical regions of increasing radius, until the enclosed density was $200$ times larger than the cosmic average density ($\langle \rho_{tot}\rangle = \langle \rho_{DM}+\rho_B\rangle$). This defines the total $M_{200}$ enclosed mass, and the virial radius $R_{200}$ assuming spherical symmetry. This procedure was implemented in a parallel routine using the Julia language\footnote{\url{https://julialang.org/}}, and it was used to analyse all ENZO simulation snapshots produced in \texttt{MAKITRA}.

\begin{figure}
    \centering
        \includegraphics[width=0.49\textwidth]{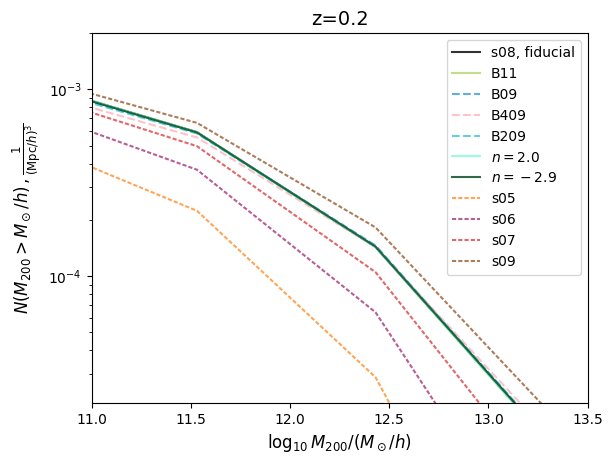}
    \caption{Halo mass distribution functions for our runs and $z=0.2$, considering the total (baryons+DM) $M_{200}$ masses. Most of the magnetic field variations (solid and dashed lines) are indistinguishable from the baseline model, unlike variations in $\sigma_8$ (dotted lines). The mass functions for runs B12, B13 and B16 are not shown for clarity, as they are indistinguishable from all others.}
    \label{fig:halo_masses}
\end{figure}

In Fig.~\ref{fig:halo_masses} we give the distribution of halo masses for all runs in \texttt{MAKITRA}, for different models at $z=0.2$. This $M_{200} \geq 10^{10} M_{\odot}$ mass range is the same we used to generate halos for the network analysis in the remainder of the paper, considering that the low mass end of the distribution is affected by the finite mass resolution of the simulation. The comparison of the mass functions clearly shows that the most important reason for variations in the \texttt{MAKITRA} suite is the change in $\sigma_8$, which directly affects the abundance of halos of all masses, as expected. Variations in the PMFs do instead produce only minor (or just invisible) differences in the mass function at $z=0.2$. Only models with $B_0 \ge 2 \rm nG$ produce barely detectable effects on the shape of the halo mass function: consequently, as we shall see, these models also produce significant differences in their network statistics.

\section{Results}
\label{sec:results}

\subsection{Halo-halo correlation functions}
We first examine the suitability of the commonly used two-point correlation function applied to our halo catalogues to discriminate between different cosmological models. 
We employ the Landy–Szalay (LS) estimator \citep{Landy:1993}:
\begin{equation}
\xi(r) = \frac{DD(r) - 2DR(r) + RR(r)}{RR(r)}\,,
\end{equation}

where $DD(r)$ is the number of data--data galaxy pairs at separation $r$, $DR(r)$ is the number of data--random pairs, and $RR(r)$ is the number of random--random pairs. All pair counts are normalised by the total number of respective pairs. We use \texttt{corrfunc}\footnote{\url{https://corrfunc.readthedocs.io}} package to compute 2ptCF.

\begin{figure*}
    
      \includegraphics[width=0.99\textwidth]{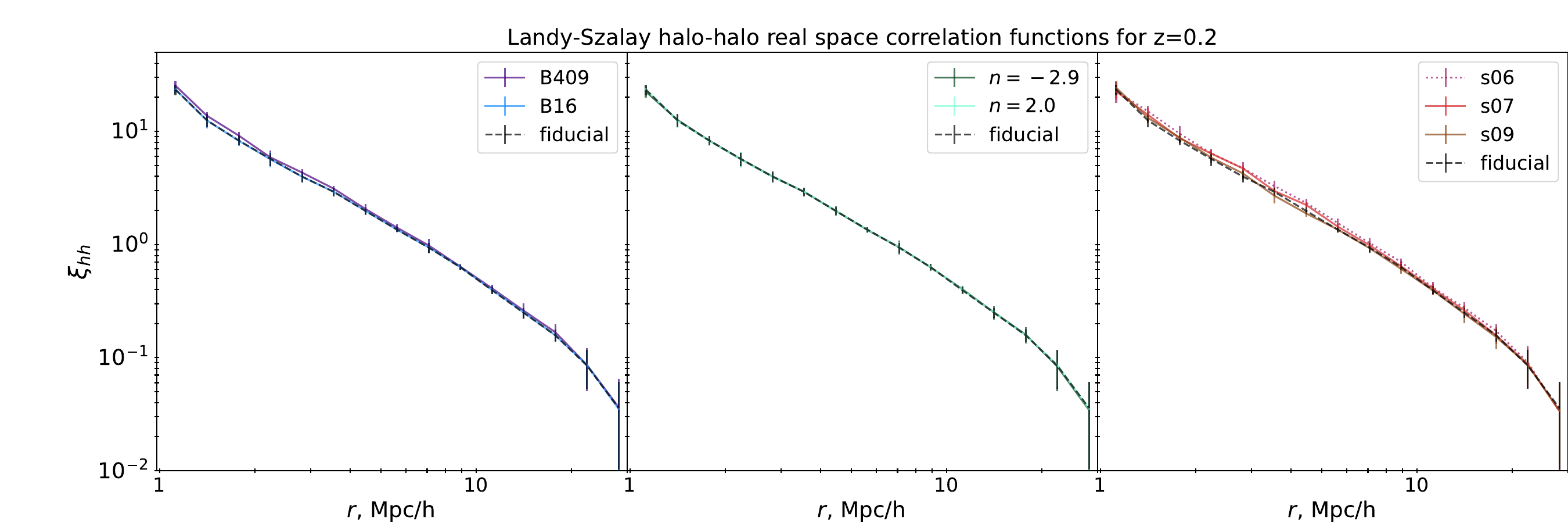}
     
     \caption{The halo-halo correlation functions $\xi_{hh}$ are computed in real space using the Landy–Szalay estimator. The error bars represent cosmic variance and correspond to the standard deviations calculated for the same distance bins across simulations with different random seeds. We show here, for clarity, only a subset of models, representing the most extreme variations of models in \texttt{MAKITRA}, to show that even in these cases, the correlation functions are basically indistinguishable.
    }
    \label{fig:ls_corr_real_space}
\end{figure*}

As anticipated in Sect.~\ref{halo_cat}, we focus only on halos with masses greater than $10^{11}\, M_\odot/h$. For the LS estimator, we use 16 logarithmically spaced bins spanning from $1\,\text{Mpc}/h$ to $31.6\,\text{Mpc}/h$. Our results are shown in Fig.~\ref{fig:ls_corr_real_space}. The correlation functions of all scenarios with variations of the magnetic field and fixed $\sigma_8$ are similar (left and middle panels of Fig.~\ref{fig:ls_corr_real_space}).  Moreover, cosmic variance does not allow us to distinguish robustly even between the $\sigma_8=0.7$ and $\sigma_8=0.8$ scenarios (right panel of Fig~\ref{fig:ls_corr_real_space}). In other words, for these simulation volumes and resolution, the variance in the halo-halo cross-correlation functions caused by the different initial conditions is larger than the difference due to the different cosmological parameters. Despite this first null test, we anticipate that the network analysis of the next Section has instead the potential of detecting cosmological variations (and to a lesser extent, PMFs variations) using the same set of halos.   

\subsection{Network metrics}
\label{network}
In this paper, we follow the logic of \cite{Tsizh:2020} and treat the cosmic web of halos as a simple undirected graph. This graph can be described in terms of the adjacency matrix $A_{ij}$-- the square matrix that describes whether the nodes $i$ and $j$ are connected by an edge or not. If nodes $i$ and $j$ are connected via edge $A_{ij}=1$, else $A_{ij}=0$. Cosmology determines the halo abundance as well as the spatial distribution; hence, it is natural to study how many neighbours the halos have and how close they are. To build a network from the halo catalogues, we use the standard \texttt{PYTHON} package \texttt{NetworkX}.\footnote{\url{https://networkx.org/}}

The structure of a network based on the point cloud strongly depends on the ``linking length'', i.e. the maximum distance between halos to consider them as connected, and it must be defined before the construction of the graph. We estimate the optimal linking length based on the percolation threshold, defined as the scale at which a giant connected component spanning a substantial fraction of the volume first emerges. For halos randomly distributed according to a Poisson process, the percolation threshold can be expressed as \citep[see, e.g.,][]{Hong:2020}:
\begin{equation}
l_{RG} = \left(\frac{3\alpha_c}{4\pi\bar{n}}\right)^{1/3},
\end{equation}
where $\alpha_c = 2.74$ in three dimensions and $\bar{n} = N/V$ is the mean number density, with $N$ the total number of halos in a volume $V$. For the cosmological scenario with $\sigma_8 = 0.8$, we obtain roughly $N \simeq 1100$ halos, so $n \approx 3.2 \cdot 10^{-4}/\rm Mpc^3$,  yielding $l_{RG} \simeq 8.5,\text{Mpc}/h$. Previous studies \citep{Hong:2020, Zhang:2018} have shown that the percolation threshold for simulated halos is slightly smaller than that for a purely random distribution. Based on this, we adopt a fiducial linking length of $8\text{Mpc}/h$, which is also qualitatively consistent with the characteristic total extent of galaxy clusters in our runs.

\begin{figure*}
    \centering
    \includegraphics[width=0.99\textwidth]{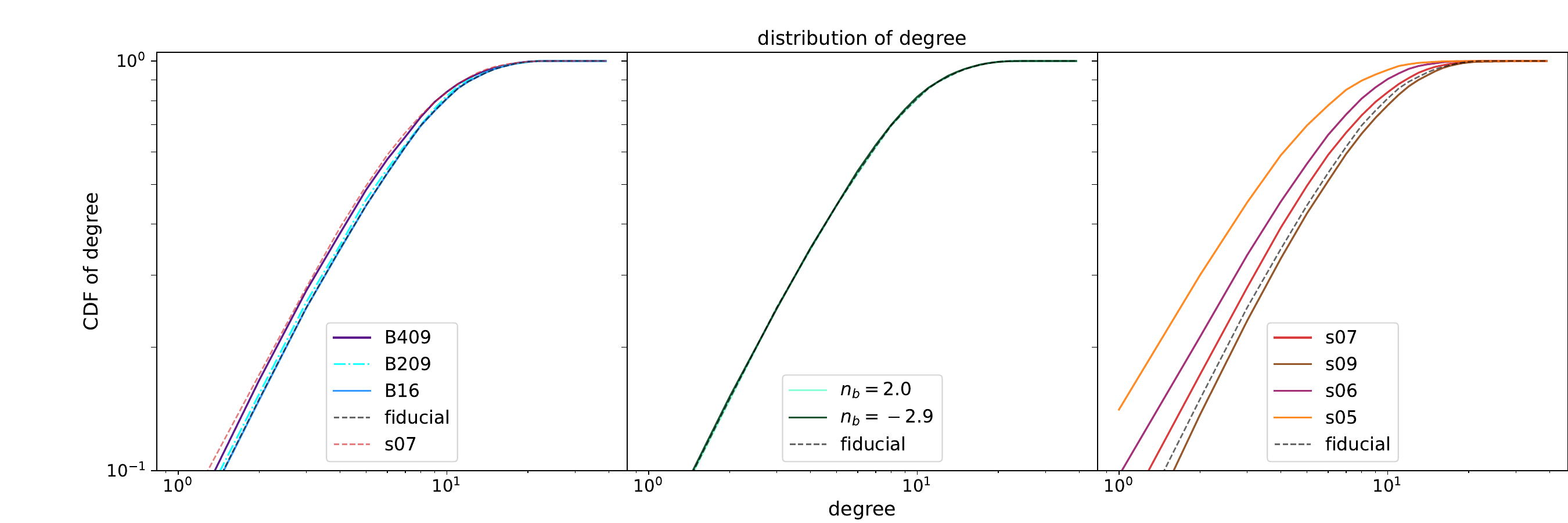}
    \caption{Cumulative distributions of the degree metric for different simulation sets at $z = 0.2$. We consider the uniform PMF scenario with a magnetic field strength of $B = 10^{-10}\,\text{G}$ and $\sigma_8=0.8$ as the fiducial (dashed line) in all panels. \textit{Left:} Uniform PMF scenarios B409, B209, B09, B16, $\sigma_8=0.7$, and the fiducial scenario. \textit{Middle:} Non-uniform PMF scenarios with power spectrum index $n$ compared to the fiducial scenario (only the two extreme spectral indices are shown for clarity). \textit{Right:} Different $\sigma_8$ scenarios with a uniform magnetic field of strength 0.1 nG compared to the fiducial scenario. }
    \label{fig:degree}
\end{figure*}

\begin{figure*}
    \centering
    \includegraphics[width=0.99\textwidth]{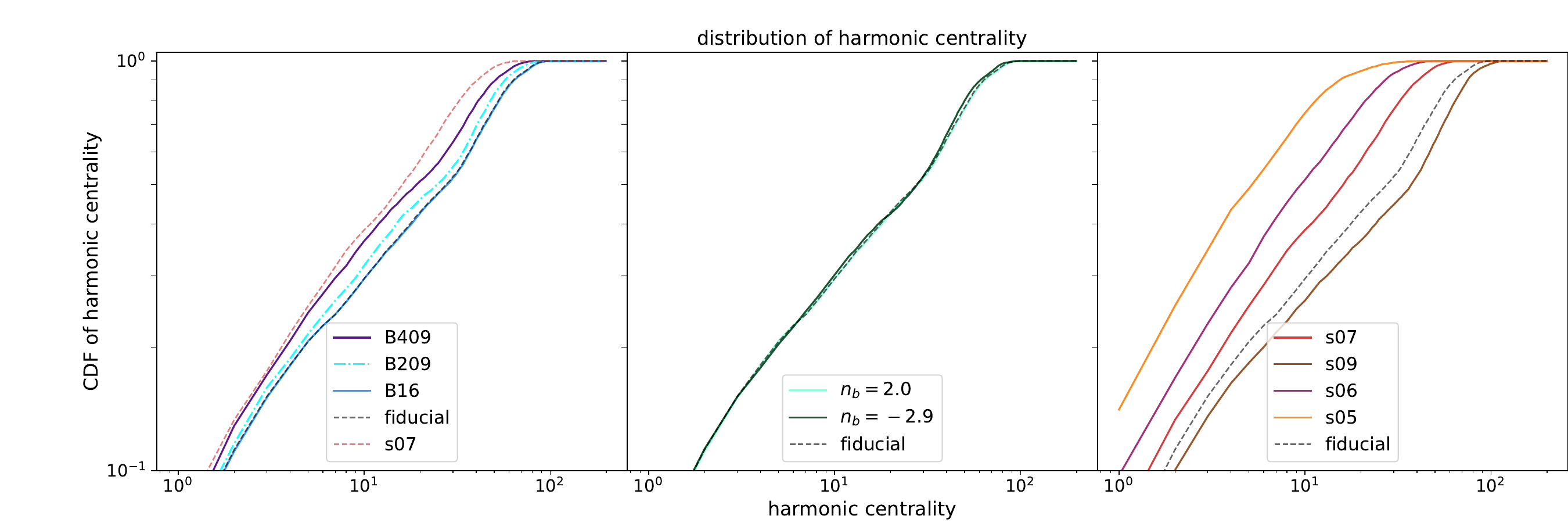}
    \caption{Cumulative distributions of the harmonic centrality for different simulation sets at $z = 0.2$ and with linking length $8~\text{Mpc}/h$. The models used in the various panels are the same as in Fig.~\ref{fig:degree}.}
    \label{fig:harmonic}
\end{figure*}

\begin{figure*}
    \centering
    \includegraphics[width=0.99\textwidth]{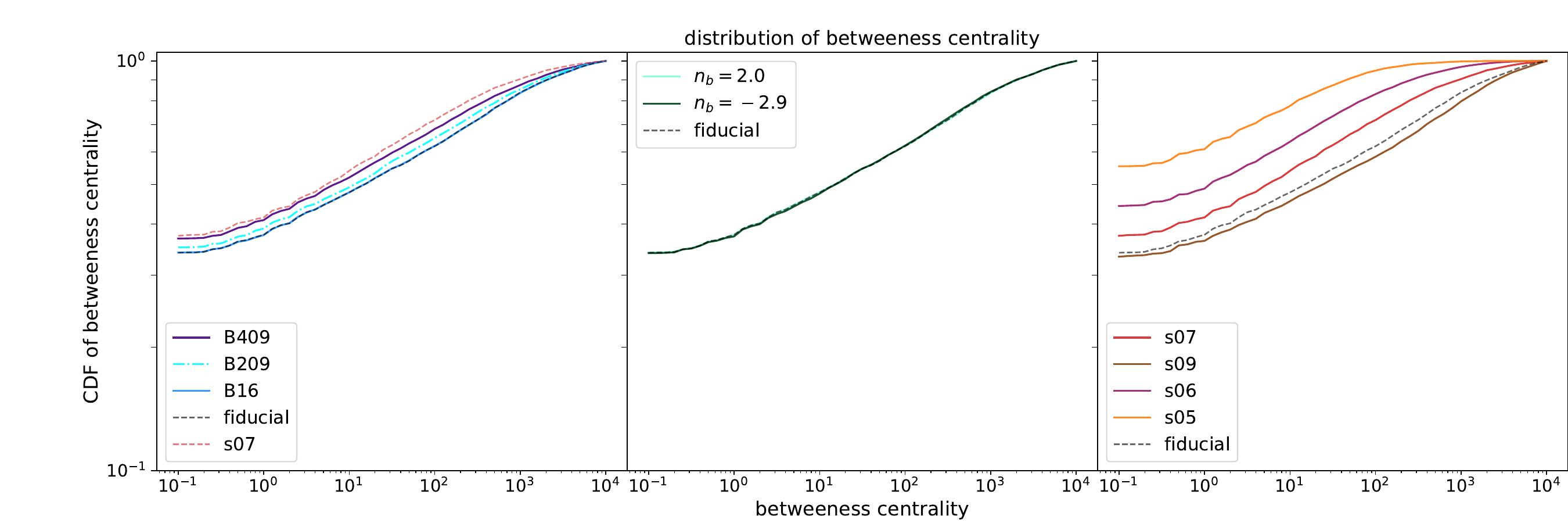}
    \caption{Cumulative distributions of the betweenness centrality for different simulation sets at $z = 0.2$ and with linking length $8~\text{Mpc}/h$. The models used in the various panels are the same as in Fig.~\ref{fig:degree}.}
    \label{fig:betweeness}
\end{figure*}

\begin{figure*}
    \centering
        \includegraphics[width=0.99\textwidth]{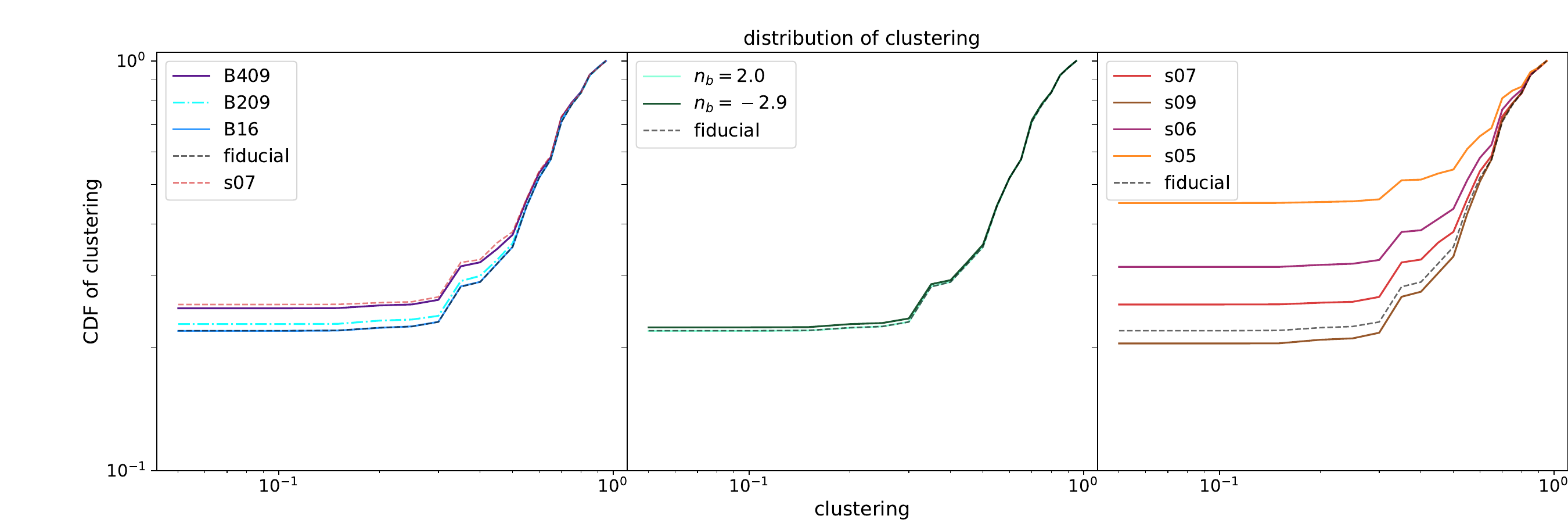}
    \caption{Cumulative distributions of the clustering for different simulation sets at $z = 0.2$ and with linking length $8~\text{Mpc}/h$. The models used in the various panels are the same as in Fig.~\ref{fig:degree}. }
    \label{fig:clustering}
\end{figure*}

To quantify the properties of networks of the simulated cosmic web in different cosmological scenarios, we exploit the normalised cumulative distribution functions, averaged over the available 8 independently simulated volumes,  for the following network metrics:
\begin{itemize}
    \renewcommand{\labelitemi}{$\circ$}

    \item the Degree $k_j$ of the node $j$, defined as:\\
    $$k_j = \sum_{i}^{N}a_{ij}\,,$$\\
    which is the number of nodes connected to this node. The degree distributions obtained for different cosmological models are presented in Fig.~\ref{fig:degree};

    \item the Betweenness centrality $C_{b}(j)$, defined as:\\

    $$C_{b}(j)=\sum^{N}_{s\neq t \neq j}{\frac{\sigma_{st}(j)}{\sigma_{st}}}\,,$$
    where $\sigma_{st}$ is is a total number of shortest paths from node $s$ to node $t$, $\sigma_{st}(j)$ is the number of those paths that connect $s$ and $t$ through $j$, and $j$ is not the end point. The betweenness centralities for our scenarios are shown in Fig.~\ref{fig:betweeness};
    
    \item the Harmonic centrality, which is the sum of the reciprocal shortest path distances from all other nodes to j:
    $$C_h(j)=\sum_{y \in V(j), y\neq j}\frac{1}{d(y,j)}\,.$$
The distributions of harmonic centrality for different models are displayed in Fig.~\ref{fig:harmonic};

\item the Clustering Coefficient, which is the ratio of the number of actual connections (edges) between its neighbours to the maximum possible number of connections that could exist among those neighbours. A high clustering coefficient indicates that a node's neighbours are more likely to be connected, forming a tightly-knit group or ``cluster'' (``clique''). The clustering coefficients for the different cosmological models are presented in Fig.~\ref{fig:clustering}.

\end{itemize}

 We find that all these network metrics are sensitive to the $\sigma_8$ variations, as expected. This produces a continuous change in the distribution of all investigated network metrics as a function of $\sigma_8$, to a degree which makes the determination of this key cosmological parameter through network analysis a very promising possibility, also considering that all these network analyses do not require a precise knowledge of the halo mass, just of their 3-dimensional position. 
When variations of PMFs are concerned, we report that only 
the extreme B409 scenario mimics the case with $\sigma_8$ decreased to $\simeq0.7$ at $z=0.2$, while the B209 is only barely distinguishable from the baseline model. This is noticeable as it contrasts with the difference in halo mass functions of these runs, as well as their correlation functions, which appear almost indistinguishable from the baseline model. This shows the potential of network analysis in detecting weak effects of PMFs correlated on very large scales, which exert a subtle dynamical effect integrated over the entire cosmic history.
All observed trends are explained with the fact that an increase of the $\sigma_8$ enhances structure formation, making the network more ``dense'' and connected, while conversely an increase of the seed field $B_0$ adds magnetic pressure to baryons and it slows down their collapse (hence roughly mimicking a decrease in $\sigma_8$), producing a larger fraction of halos "isolated" from the rest of the network. 
 Nevertheless, all other investigated primordial magnetic field variations and $\sigma_8=0.8$ appear to be barely distinguishable from the baseline scenario, also via network metrics. 

We further elaborate on the physical reason for this in Sect.~\ref{sec:discussion}.

\subsubsection{Network centralities: cosmic variance impact}
\begin{figure*}
    \centering
        \includegraphics[width=1.0\textwidth]{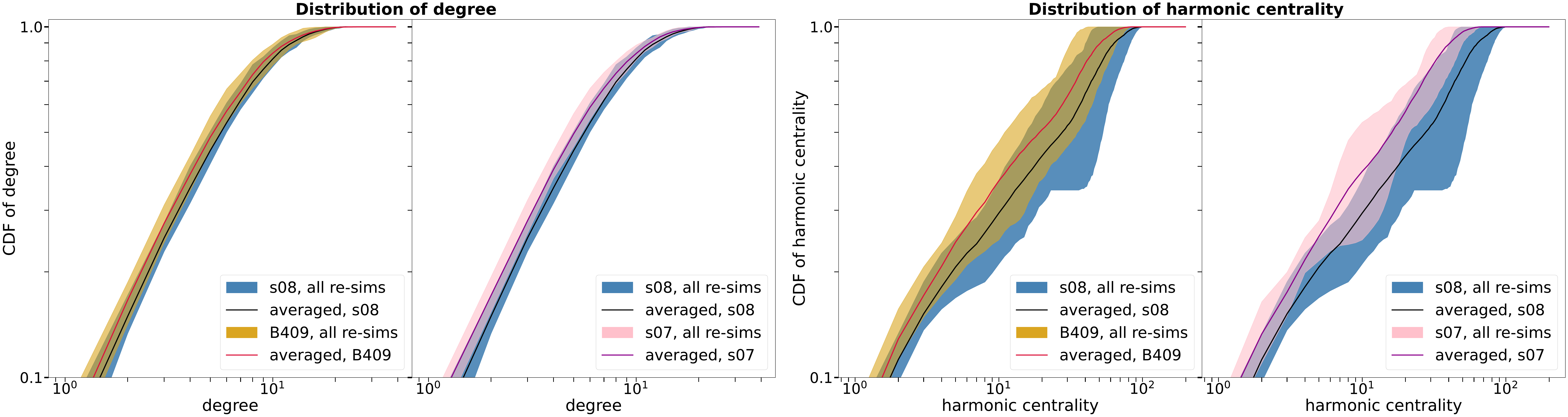}
    \caption{Cumulative distributions of the degree (left panel) and harmonic centrality (right panel) for the B409, s07, and fiducial scenarios. Solid lines represent the CDF averaged over different re-simulations, while shaded regions indicate the variance of the CDFs due to different random initial conditions (cosmic variance).}
    \label{fig:CV_degree_harmonic}
\end{figure*}

\begin{figure*}
    \centering
        \includegraphics[width=1.\textwidth]{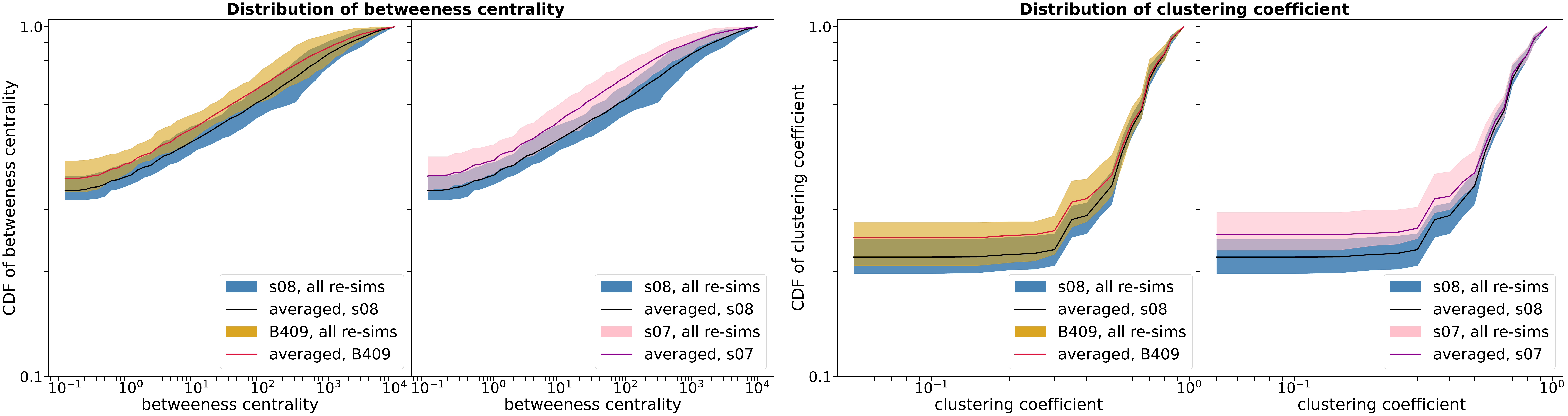}
    \caption{The same as in the previous Figure, but for the betweenness and clustering centralities.}
    \label{fig:CV_betweeness_clustering}
\end{figure*}

In the previous subsection, we focused on the distributions of node centralities, averaged over multiple re-simulations. Nevertheless, cosmic variance, arising from different random seeds,  remains significant even for simulations that in total cover $300^3\ \rm Mpc^3$. To assess the role of cosmic variance, 
we show the cumulative distributions of the four aforementioned centralities across all available re-simulations, comparing B409 and s07 models against the fiducial scenario. The results are summarized in Figs.~\ref{fig:CV_degree_harmonic} and~\ref{fig:CV_betweeness_clustering}.

We find that the empirical scatter between distributions of degree, harmonic centrality, betweenness, and clustering centrality across the different re-simulations of each model is comparable with the relative difference produced by variations of the physical models. On the other hand, the difference between the averaged CDFs is less than the variance of the scatter of CDFs across re-simulations of the same models. This fact demonstrates the potential of network analysis to overcome the problem of cosmic variance.

By assuming a standard Poissonian statistics, scaling with the square root of the number of halos used to reconstruct the network, we estimate that in order to disentangle variations of $\sigma_8$ of order of $0.1$, or differences in $B_0$ of order of 2 (for $B_0\geq 2 \rm nG$) a total volume of $\rm Gpc^3$ would be enough to overcome the effect of cosmic variance, based on the trend measured with our simulations.

\subsection{Counts-in-cell statistics}

The previous tests showed that the maximally magnetised scenario B409, as well as cosmological models with $\sigma_8 \leq 0.7$, produce similar network centralities. To further investigate why, we apply counts-in-cell statistics to our simulations. The CIC statistic shows the expected number of halos that can be found in a randomly placed sphere with radius $R$. We exploit the \texttt{corrfunc} package to compute CIC.

\begin{figure*}
    
      \includegraphics[width=0.99\textwidth]{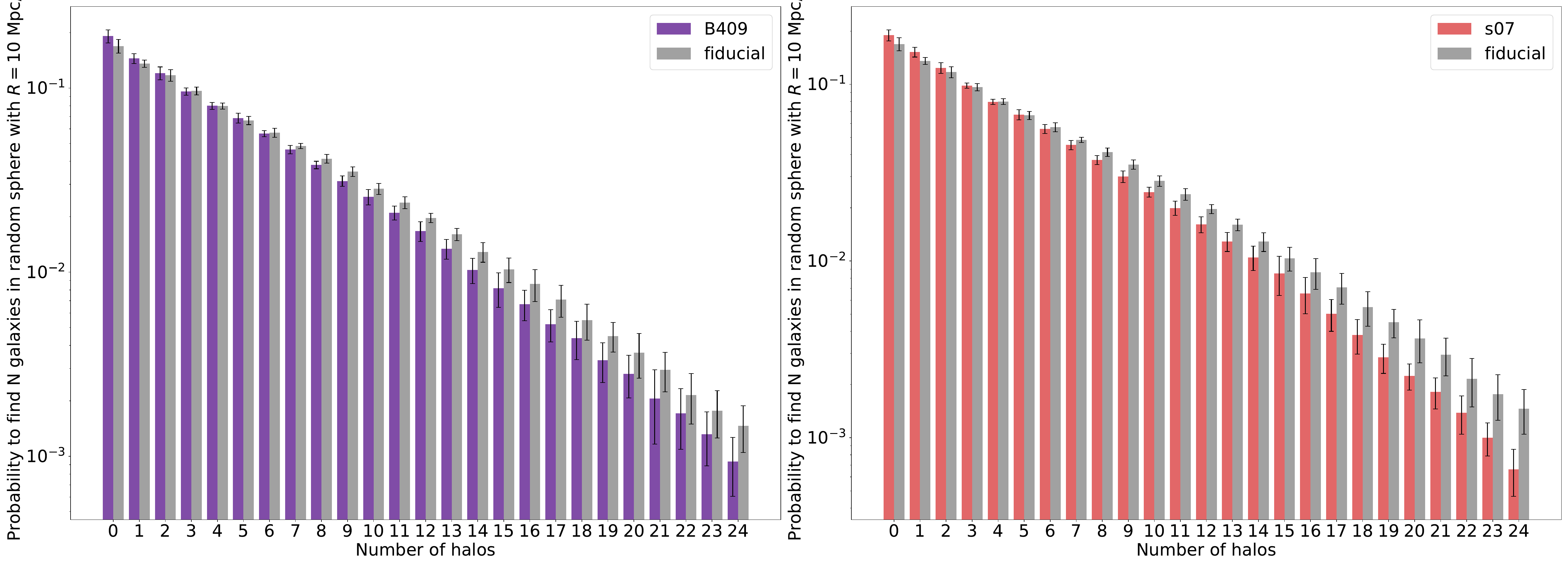}
     
    \caption{The counts-in-cell statistic computed for fiducial, B409 and s07 models for $z=0.2$. The sphere radius is chosen to be $R=10\,\text{Mpc}/h$. Errors correspond to the scatter (in terms of sample standard deviation) of the probability to find $N$ galaxies in a sphere among the different simulation realisations.}
    \label{fig:CIC_fig}
\end{figure*}

Our results for this statistic are provided in Fig.~\ref{fig:CIC_fig}.  The CIC statistics confirm the indications of network centralities:  both B409 and s07 runs produce a larger fraction of ``isolated'' halos (defined as halos whose nearest neighbours lie at distances greater than a chosen threshold, e.g., $8\,\mathrm{Mpc}$) compared to the baseline model. 

We should mention that the probability of finding $N$ galaxies in the sphere with radius $R$ estimated for some finite volume can be treated as a random variable because of the cosmic variance. Thereby, for a given sample of the simulations with different random initial conditions, we can compute the sample characteristics, e.g., ``mean'' probability to find $N$ galaxies in a sphere with radius $R$, as well as standard deviations, computed across the considered re-simulations. 

While both B409 and s07 models demonstrate a lower probability of finding some number of halos in the sphere with radius $R=10\,\mathrm{Mpc/h}$ compared to the fiducial ($\sigma_8=0.8$) scenario, the cosmic variance makes attempts to distinguish between different cosmological models challenging.

\section{Discussion}
\label{sec:discussion}

In this work, we have presented the first application of network analysis to a suite of cosmological magneto-hydrodynamical simulations, aimed at assessing the potential of this approach to constrain cosmological models. Our findings indicate that network-based statistics have the capacity to constrain the cosmological parameter $\sigma_8$. 

We note that we did not investigate the ability of network centrality statistics to break the $\sigma_8$–$\Omega_m$ degeneracy. Previous studies have shown that incorporating higher-order galaxy correlations (e.g., bispectrum) into parameter inference can significantly tighten constraints on these cosmological parameters \citep[see, e.g.][]{Sefusatti:06, Hahn:2020}. Since network centralities naturally encode information from correlations beyond the two-point level, they represent a promising tool for improving joint constraints on both $\sigma_8$ and $\Omega_m$. A detailed assessment of the joint constraining power of network analysis is left for future work. Such an investigation could be carried out, for instance, by combining machine-learning techniques with network-based statistics and a large suite of simulations that vary both $\Omega_m$ and $\sigma_8$, such as the \texttt{CAMELS-SAM} dataset \citep{Perez:2022nlv}. 
 
Our study has reported that a higher value of $\sigma_8$ enhances structure formation and increases clustering, producing a more interconnected network with a larger number of halo–halo links. This result is consistent with the findings of \citet{Tsizh23}, who demonstrated that topological analyses applied to the cosmic web, particularly through persistence homology, provide a powerful tool for constraining $\sigma_8$.  
An important advantage of the network analysis is worth stressing once more:  it only makes use of the 3-dimensional position of matter halos, while an estimate of their mass is not needed (however, the definition of a lower threshold value for the mass is needed in our procedure, see Sect.~\ref{network}). This makes it an extremely convenient tool to constrain cosmology with modern survey data. 
Furthermore, \citet{Hong:2020} argued that degree centrality of a cosmological graph naturally encodes information from higher-order correlations beyond the traditional two-point statistics, providing an additional theoretical motivation for using network centralities as probes of cosmological models. 
On the other hand, our analysis exposed the limitations introduced by cosmic variance, in the sense that our results demonstrate a significant impact of cosmic variance on the network centralities statistic. 
In this paper, we exploit the hydrodynamical simulations with volume $(150\,\text{Mpc})^3=(101.7\text{Mpc}/h)^3$, which were re-simulated 8 times within different random seeds, for a total $~(300\,\text{Mpc})^3$ volume. This volume is in the same order as the existing SDSS survey, yet it is much smaller than the expected Euclid data,  \cite[see, e.g.][]{Euclid:2024yrr}.
Though the convergence of statistics of network centralities with volume was not studied, we estimated the needed volume to reduce the effect of the cosmic variance below the typical separation between models in the network statistics, suggesting that a $\sim \rm Gpc^3$ volume should be the minimum needed.
As a comparison, \citet{Schaller:2025} reported that 
the total matter power spectra at $z=0$ is converged for hydrodynamical simulations with volumes $\gtrsim(100\,\text{Mpc}/h)^3$, while  \citep{Hong:2020} found that size $(256\,\text{Mpc}/h)^3)$ subvolumes is not large enough to suppress the cosmic variance effects on the statistics of large connected components,  as well as on the graph transitivies of the cosmic graphs computed for the $N$-body simulations.
To be conservative, as a result of our work and also including the effects of baryonic feedback (which we only explored in the Appendix), we conclude that a $(0.5-1\,\text{Gpc})^3$ volume is the minimum to reach the convergence of the network statistics. 

The natural question is: are the existing and upcoming surveys data accurate enough to be exploited for the network analysis? In the near future, SKA-Mid should be able to perform an unprecedentedly precise survey of galaxies in the cosmic web. It is expected to measure redshifts with an accuracy of $\Delta z = 1-5\cdot10^{-5}$ and its angular resolution will reach approximately $0.5\,\text{arcsec}$ \citep[see, e.g.][]{Braun:19, Hartley:23}. 
For the redshift $z=0.2$ considered in this work, these specifications correspond to positional uncertainties on the order of a few to tens of kiloparsecs, which is significantly smaller than our fiducial linking length  $8\,\text{Mpc}/h$.

Among already existing surveys, the Sloan Digital Sky Survey (SDSS) provides angular accuracies of order $\sim 0.1,\text{arcsec}$ and radial-velocity uncertainties of tens of kilometres per second \citep{Bolton:2012}, corresponding to $\Delta z \sim 10^{-4}$. Consequently, and similarly to the previous case, the observational uncertainties in the SDSS catalogues are smaller than our chosen linking length and therefore should not significantly affect the construction of the network. Considering the ongoing survey with the Euclid satellite, its redshift accuracy in tomographic bins is anticipated to be $\approx 0.002$ \citep{EuclAcc}.
We thus conclude that network-based analysis can be robustly applied to both current and future galaxy surveys, provided that the entire scanned volume is of order $\sim \rm Gpc^3$, as motivated above. 

Another aspect to notice is that galaxy surveys measure redshifts, which are affected by peculiar velocities. Consequently, using observables measured in the so-called redshift space is more realistic than working in real space. Moreover, the use of redshift-space measurements can help to break the $\Omega_m$–$\sigma_8$ degeneracy, as has been demonstrated for galaxy correlation functions \citep[see, e.g.,][]{Tinker:2006}. We repeated our analysis in redshift space and found that network-centrality statistics remain equally sensitive to variations in $\sigma_8$. A comprehensive investigation of the constraining power of network analysis applied to cosmic graphs in redshift space is beyond the scope of this paper and is also deferred to future work.

Concerning the additional potential impact of primordial magnetic fields on the network centralities, we found that the cosmological scenario with large primordial uniform values ($B_0 = 2-4\,$nG) produces more isolated massive halos ($\geq 10^{11}\, M_\odot/h$) than all other less magnetised models with the same $\sigma_8$(=0.8). This stems from the volume-filling extra pressure from magnetic fields, which slows down the collapse of baryonic matter into halos and prevents the formation of a significant fraction of small mass halos.

\begin{figure}
          \includegraphics[width=0.45\textwidth]{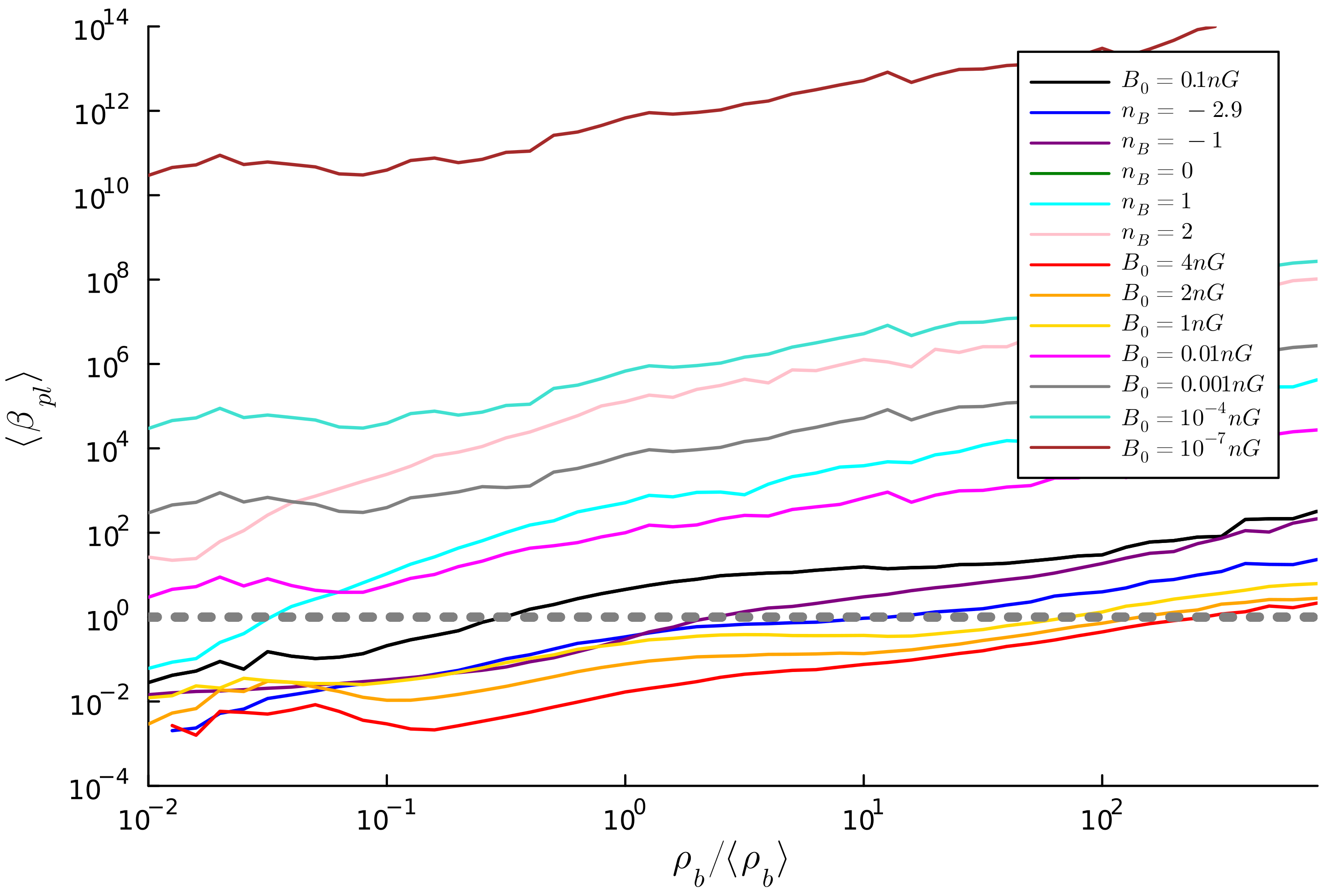}
    \caption{Distribution of the average plasma beta, $\beta_{pl}$, as a function of the ordinary matter over density, for the same runs of Fig.\ref{fig:PDF_B} (the horizontal line marks $\beta_{pl}=1$).}
    \label{fig:PDF_beta}
\end{figure}

In general, the role of PMFs in affecting the evolution of small-mass halos has been reported by previous work already: PMFs were also shown to enhance the formation of low-mass ($10^8-10^9\, M_\odot$) halos via amplification of the matter power spectrum at small scales, while magnetic pressure suppresses the growth of smaller scale density perturbation by effectively increasing their Jeans length \citep[see, e.g..][]{Kim:94, Kahniashvili:2012dy, Sanati:20, Cruz:24}. 
However, the network statistics discussed in our work show a proportional dependence on $\sigma_8$, but a discontinuous dependence on the amplitude of PMFs: only for $B_0 \simeq 4\,\rm nG$ seed fields the network statistics show some difference from the baseline model, in a rather abrupt transition. 
Why is that?  This can be understood by noticing that the thermal gas pressure in most of the cosmic volume is set to a floor value by the realistic UV re-heating background from quasars added to the simulation at $z \sim 7$, following \citet{HM12}. After reionization, the mean comoving baryon density is $\bar{n} \approx 2.5\times 10^{-7}\,\mathrm{cm^{-3}}$  \citep{page2003first}. For a reference $T \sim 10^4 \rm K$ temperature re-heating floor, the average baryon pressure energy density is $p_{th}=n k_B T/ \sim 2.7 \cdot 10^{-19} \rm erg/cm^3$ (where $n$ is the baryon number density and $T$ is the baryon temperature), while the average magnetic pressure energy density is  $p_B=B^2/8\pi\sim 6.4 \cdot 10^{-19} \rm erg/cm^3$ for $B_0=4 \rm nG$, and $p_B\sim 1.6 \cdot 10^{-19} \rm erg/cm^3$ for $B_0=2 \rm nG$. For all smaller magnetic field amplitudes, it is $p_B \ll p_{th}$, and thus from the dynamical viewpoint, all networks behave in the same way, regardless of their magnetic pressure. This derivation is only approximate: on one hand, the UV re-heating temperature evolves with redshift, and on the other, the magnetic field amplitude depends on cosmic overdensity. 
A more quantitative view of the dynamical balance between thermal and magnetic pressure is given by the distribution of the plasma beta parameter ($\beta_
{pl}=p_{th}/p_B$) for the same models of Fig.~\ref{fig:PDF_B}. $\beta_{pl}$ parametrises the importance of thermal gas pressure versus magnetic pressure, and its average value as a function of the cosmic matter overdensity, for all runs, is given in  Fig.~\ref{fig:PDF_beta}. 
As expected, runs with the highest $B_0$ have $\beta_{pl} \leq 1$ below the cosmic overdensity of filaments and voids, which refers to a very large fraction ($\geq 90 \%$) of the cosmic volume: there magnetic pressure is so significant that it has a measurable effect on the clustering of matter even at the $z=0.2$. In this sense, large PMFs tend to mimic universes with low $\sigma_8$ in the production of more isolated halos in voids. While in simulated universes with a larger $\sigma_8$, halos are affected by the accelerated matter growth for the entire duration of the cosmic evolution, the effect of different initial $B_0$ is mostly erased (for $B_0 \leq \rm 1~nG$ fields) by the generation of a temperature floor for $z \leq 7$, due to UV radiation. Although the details of the effect may depend on the adoption of a specific UV reheating background, the basic physics of this effect is robust, and we conclude that it should always be considered when assessing the dynamical effect of primordial magnetic fields. 

Finally, we notice that our simulations do not include all plausible effects of PMFs in the initial conditions, like the sourcing of small-scale matter and velocity fluctuations. Furthermore, we did not delve into the possible effects of the baryonic feedback on the statistics of the centralities of the cosmic networks. 
The baryonic feedback can significantly impact the small-scale matter power spectra \citep[e.g.][and references therein]{Schaller:2025}. Our preliminary results highlight that the stellar and active-galaxy nuclei can significantly impact the cosmic network, see more in the Appendix~\ref{sec:appendix_feedback}. We leave the detailed simulation of these effects and the application of the network analysis to such simulations for future studies.

\section{Conclusions}
\label{sec:conclusions}

   Using a new suite of cosmological hydrodynamical simulations,  we tested the sensitivity of network statistics as a tool to identify subtle variations in cosmological parameters  (namely the $\sigma_8$ parameter) or in the properties of primordial magnetic fields (i.e. either by varying their initial amplitude and/or spectral slope).   The impact of PMFs on the growth of large structures is potentially important for understanding both the origin of cosmic magnetism and for allowing a precise reconstruction of cosmological parameters from existing and future galaxy survey data.  
   
   We have analysed several types of metrics characterizing the spatial distribution of dark and baryonic matter halos: network centralities' statistics, two-point correlation functions, and counts-in-cell statistics. We considered halos as the vertices of the Cosmic web's graph and computed several network metrics for each vertex. Our main results can be summarised as follows:

   \begin{itemize}
   \item  The statistics of the network centralities can serve as a novel sensitive probe of the cosmological parameter $\sigma_8$. Variations of $\Delta \sigma_8=0.1$  can be robustly detected by comparing all distributions of network parameters at $z=0.2$. 
   \item The typical level of cosmic variance in the halo distribution measured in our simulation suggests that, in order to constrain $\sigma_8$ to a $\Delta \sigma_8 \leq 0.1$ level through network statistics, it should be necessary to analyse a $\geq \rm Gpc^3$ cosmic volume. 
   \item  The network centralities allow us to distinguish cosmological scenarios with variations of PMF's strength, even when their halo mass functions are seemingly identical. Large PMFs affect the network statistics in a way which broadly mimics a $\sigma_8 \approx 0.7$ cosmology. However, this is limited to $B_0\geq 2\,\mathrm{nG}$ seed fields, which are at the edge (of slightly beyond) the upper limits which can be derived from CMB analysis at the epoch of recombination \citep[e.g.][]{1997PhRvL..78.3610B,2019JCAP...11..028P}.
   \item    PMFs with an effective lower normalisation are virtually indistinguishable from the baseline model. The presence of a realistic UV re-heating background, introduced at $z \sim 7$ in the simulation, indeed increases the thermal pressure of baryons everywhere and almost completely overcomes any dynamical effect of PMFs on the dynamics of the matter network.
   \end{itemize}

Finally, we stress the important aspect that network analysis is a potentially very powerful tool for cosmology, since the exact mass distribution of halos (provided they are $M_{200}\geq 10^{11}\,M_{\odot}/h$) is not required for the network analysis, just a reconstruction of their 3-dimensional location. 
As existing and upcoming surveys map the galaxy distribution deeper in redshift and with high precision, these methods provide a valuable complement to traditional clustering statistics, towards a full characterisation of all matter and energy components of the Universe.

\begin{acknowledgements}
    AR has been partially supported by the Scholars at Risk Fellowship from the Physics \& Astronomy Department of the University of Bologna. FV has been partially supported by Fondazione Cariplo and Fondazione CDP, through grant n$^\circ$ Rif: 2022-2088 CUP J33C22004310003 for the ``BREAKTHRU'' project.  FV acknowledges the CINECA award  ``IsCc4{\_}FINRADG" and  ``IscrB{\_}CREW"  under the ISCRA initiative, for the availability of high-performance computing resources and support, and the usage of online storage tools kindly provided by the INAF Astronomical Archive (IA2) initiative (http://www.ia2.inaf.it).  F.V. also acknowledges the usage of computing at the Gauss Centre for Supercomputing e.V. (www.gauss-centre.eu) for supporting this project by providing computing time through the John von Neumann Institute for Computing (NIC) on the GCS Supercomputer JUWELS at J\"ulich Supercomputing Centre (JSC), under the project ``BREAKTHRU''. MT was supported by the Ministry of Education and Science of Ukraine, with grant number 0121U113567. This work was also partially supported by the NAS of Ukraine under project number 0121U109612.
\end{acknowledgements}

\bibliographystyle{aa} 
\bibliography{mybib,franco3} 

 \appendix

\section{The \texttt{MAKITRA} Suite}
\label{A1}

\texttt{MAKITRA} is a general-purpose suite of cosmological 
 ideal MHD simulations produced with the Eulerian \texttt{ENZO} code, in a fixed-grid modality \cite[see, e.g.][]{Enzo:2014}, covering 8 independent volumes of $150^3 \rm Mpc^3$, therefore sampling a total $300^3 \rm Mpc^3$ cosmic volume.  Several of its data products are publicly available at \url{https://cosmosimfrazza.eu/makitra}. 
 \texttt{MAKITRA}  is meant to investigate variations of cosmological, baryonic feedback, and magnetic parameters, and in this paper, we analysed a large subset of its model variations.  
 
The full list of parameter variations and runs of the \texttt{MAKITRA} suite (also including those not used in this paper, at the end of the Table) is given in Table \ref{tab_cosmo}.

\begin{figure*}
\centering
\includegraphics[width=0.95\textwidth]{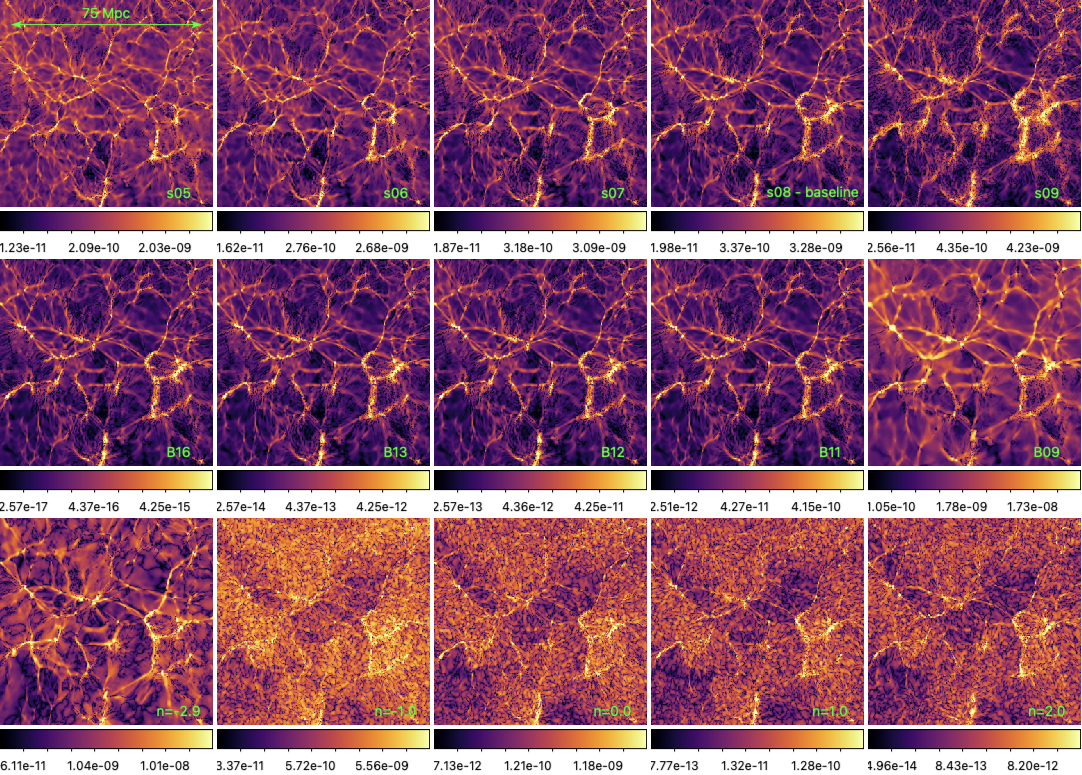}
\caption{Maps of magnetic fields strength, in units of [G], for a thin slice ($292$ kpc thick along the line of sight) crossing one of our simulated boxes, at $z=0.2$, for all model variations discussed in the paper. The first row shows variations of $\sigma_8$, the second shows variations in the initial amplitude of the uniform magnetic field seed $B_0$, and the last row shows variations in the slope $n$ and normalisation of the initial magnetic spectra.  The range of values covered by each colorbar has been adjusted to match the range of values of each run.} 
\label{fig:new_sim}
\end{figure*}

 Fig.~\ref{fig:new_sim} shows the distribution of magnetic field amplitude, for a thin slice through one of our simulations at $z=0.2$, for 8 key variations of the magnetic field scenario considered in this work, including variations of amplitude or initial power spectrum index of the primordial fields. This clearly shows the different volume filling and 3-dimensional organisation of the fields, which all wrap around the same underlying cosmic web structures, shaped by dark and baryonic matter.

\begin{table*}
\label{tab_cosmo}
\caption{List of all cosmological ENZO runs in the  \texttt{MAKITRA}suite, with their main parameters.}
\centering \tabcolsep 5pt 
\begin{tabular}{c|c|c|c|c|c|c|c|c|c|c|}
Run ID & $L_{box}$  & $\Delta x$ & $\Omega_M$ &  $\Omega_b$  &  $\Omega_\Lambda$ & $\sigma_8$ & $h$ & $n$ &$\langle B_0 \rangle_{\rm Mpc}$ & physics \\
       &   [Mpc]    &  [kpc]      &       &      &     &     & [$100 ~\rm km/s/Mpc$] & & [G] & \\ \hline
s08 (Baseline)& 150  &  293 &  0.308 & 0.0478 & 0.692 & 0.8& 0.678 &  -     & $10^{-10}$& cool+UV+MHD\\
 s05& 150  & 293 & 0.308 & 0.0478 & 0.692 & 0.5& 0.678 & -     & $10^{-10}$&cool+UV+MHD\\
 s06& 150  & 293 & 0.308 & 0.0478 & 0.692 & 0.6& 0.678 & -     & $10^{-10}$&cool+UV+MHD\\
 s07& 150  & 293 & 0.308 & 0.0478 & 0.692 & 0.7& 0.678 & -     & $10^{-10}$&cool+UV+MHD\\
 s09& 150  & 293 & 0.308 & 0.0478 & 0.692 & 0.9& 0.678 & -     & $10^{-10}$&cool+UV+MHD\\
 s10& 150  & 293 & 0.308 & 0.0478 & 0.692 & 1.0& 0.678 & -     & $10^{-10}$&cool+UV+MHD\\
  B11& 150  &  293 &  0.308 & 0.0478 & 0.692 & 0.8& 0.678 & -     & $10^{-11}$& cool+UV+MHD\\
   B16& 150  & 293 & 0.308 & 0.0478 & 0.692 & 0.8& 0.678 & -     & $10^{-16}$&cool+UV+MHD\\
   B13& 150  &  293 &  0.308 & 0.0478 & 0.692 & 0.8& 0.678 & -     & $10^{-13}$& cool+UV+MHD\\ 
   B12& 150  &  293 &  0.308 & 0.0478 & 0.692 & 0.8& 0.678 & -     & $10^{-12}$& cool+UV+MHD\\
    B09& 150  &  293 &  0.308 & 0.0478 & 0.692 & 0.8& 0.678 & -     & $10^{-9}$& cool+UV+MHD\\
    B209& 150  &  293 &  0.308 & 0.0478 & 0.692 & 0.8& 0.678 & -     & $2 \cdot 10^{-9}$& cool+UV+MHD\\
    B409& 150  &  293 &  0.308 & 0.0478 & 0.692 & 0.8& 0.678 & -     & $4 \cdot 10^{-9}$& cool+UV+MHD\\
 $n=-2.9$& 150  & 293 & 0.308 & 0.0478 & 0.692 & 0.8& 0.678 & -2.9& $2.0 \cdot 10^{-9}$&cool+UV+MHD\\
 $n=-1$& 150  & 293 & 0.308 & 0.0478 & 0.692 & 0.8& 0.678 & -1.0& $1.8 \cdot 10^{-9}$&cool+UV+MHD\\
 $n=0$& 150  & 293 & 0.308 & 0.0478 & 0.692 & 0.8& 0.678 & 0.0& $3.5 \cdot 10^{-10}$&cool+UV+MHD\\
 $n=1$& 150  & 293 & 0.308 & 0.0478 & 0.692 & 0.8& 0.678 & 1.0& $4.2 \cdot 10^{-11}$&cool+UV+MHD\\
 $n=2$& 150  & 293 & 0.308 & 0.0478 & 0.692 & 0.8& 0.678 & 2.0& $3.0 \cdot 10^{-12}$&cool+UV+MHD\\
 B16\_feedb1& 150  & 293 & 0.308 & 0.0478 & 0.692 & 0.8& 0.678 & -     & $10^{-16}$&cool+UV+MHD+star\\
 B16\_feedb2& 150  & 293 & 0.308 & 0.0478 & 0.692 & 0.8& 0.678 & -     & $10^{-16}$&cool+UV+MHD+AGN \\
  B16\_feedb3& 150  & 293 & 0.308 & 0.0478 & 0.692 & 0.8& 0.678 & -     & $10^{-16}$&cool+UV+MHD+star/AGN \\
 
\end{tabular}
\end{table*}

 \section{Modelling of primordial magnetic fields}
\label{A2}
To introduce a 3-dimensional magnetic field model in all primordial scenarios, we followed previous work \citep[][]{va21} and modelled primordial magnetic fields as a stochastic background, constrained by the two-point correlation function \citep[e.g.][]{Finelli:2008xh}:
\begin{equation}
\langle B^\star_i({\mathbf k})B_j({\mathbf k'})\rangle =
\delta^{(3)} ({\mathbf k}-{\mathbf k'}) P_{ij}({\mathbf{\hat k}})
P_B(k) (2\pi)^3,
\end{equation}
where $i$ and $j$ are spatial indices. $\delta^{(3)}({\mathbf k}-{\mathbf k'})$ is the Dirac delta function, with unit vector 
$\hat{k}_i=k_i/k$, $P_{ij}({\mathbf{\hat
k}})=\delta_{ij}-\hat{k}_i\hat{k}_j$ is the operator for transverse plane projection, and $P_B(k)$ is the power spectrum of the magnetic field.
The (inverse) scale dependence of the field is prescribed by a power law spectrum: $P_B(k) = P_{B0}k^{n} $, where $n$ is the spectral index. Following the convention in the literature, we describe the amplitude by smoothing the fields on the scale $\lambda=1 \rm ~Mpc$ comoving, and using $B_\lambda$:
\begin{equation}
P_B(k) = P_{0}k^{n}= \frac{2\pi^2 \lambda^3
B^2_\lambda}{\Gamma(n/2+3/2)} (\lambda k)^{n}.
\label{energy-spectrum-H}
\end{equation}
For $k>k_D$, where 
$k_D$ is the cutoff wavenumber, beyond which the magnetic energy is supposed to be dissipated through  Alfv\'en wave damping, on scales smaller than the Silk damping scale for acoustic waves, with 
alternating phases of turbulent and viscous regimes \citep[e.g.][]{2004PhRvD..70l3003B}. It should be noted that the physics ruling the exact value of $k_D$ is complex \citep[e.g][]{Trivedi18, Jedamzik_Abel_23} as it depends on the combination of $\alpha_B$ and $B_\lambda$. However, the spatial scale $\sim 2\pi/k_D$ is in the $\sim 10 \rm ~kpc$ range, which is smaller than our grid resolution, and hence the effects of $k_D$ on our results can be considered, to a first approximation, not large. 
We assumed a range of different cases for $n$, from the almost scale-invariant $n=-2.9$ (whose energy distribution in Fourier space is infrared dominated), to the high value of $n=2$ (which corresponds to the minimum index allowed for causally, i.e., post-inflation, generated magnetic fields).  
The values for the amplitudes of the fields are constrained 
by the analysis of Cosmic Microwave Background multipoles \citep[][]{2019JCAP...11..028P} For the homogeneous field case, we tested a range of values from $4 \cdot 10^{-9} ~\rm nG$ (of the order of the  COBE constraints \citealt[][]{1997PhRvL..78.3610B}, obtained however for a different cosmology), to $10^{-16} \rm nG$, which is at the level of existing lower limits from $\gamma$-ray analysis of blazars \citep[e.g.][and references therein]{2021Univ....7..223A} and it effectively represents a non-magnetised case in our analysis as magnetic fields of this amplitude cannot exert any pressure on the gas. 
It shall be noticed that the fiducial cosmological parameters are not varied for the different magnetic fields model,  because their inclusion does not introduce strong degeneracies with the standard cosmological parameters \citep[e.g.][]{PLANCK2015,2019JCAP...11..028P}. Several studies have argued that primordial magnetic fields can exert an impact on the matter and velocity power spectrum of both the baryonic and dark matter components. 
\citep[e.g.][]{Sethi:2004pe, Finelli:2008xh, Fedeli:2012rr, 2013ApJ...770...47K, Ralegankar:2024}, although the exact details of the perturbation to consider of the standard (i.e., un-magnetised) cosmological model are non-trivial to compute \citep[e.g.][]{Pavicevic:2025}.

In this case, the initial magnetic field configurations are imposed at the start of runs, $z_{\rm ini}= 30$, hence magnetic fields couple to baryonic matter only at this stage. The elapsed time between recombination and $z_{\rm in}$ is $\approx 60 \rm ~Myr$, which is very short compared to the total time of the simulation, and no significant structure formation or evolution of magnetic fields is expected. 

Finally, for simplicity in this work, we only consider non-helical magnetic field configurations. Inflationary generation processes can produce helical magnetic fields 
\citep[e.g.][]{Field:1998hi, Vachaspati:2001nb, Sigl:2002kt, Sharma:2018kgs}, with a potentially relevant impact on on CMB anisotropies  \citep[e.g.][]{Pogosian:2002dq, Caprini:2003vc, Ballardini:2014jta, Kahniashvili:2014dfa, Zucca:2016iur}. The cosmological simulation of helical primordial fields with ideal MHD simulations has only recently begun \citep[][]{2022ApJ...929..127M} and might be considered in future work.

\section{The baryonic feedback impact on cosmic networks }\label{sec:appendix_feedback}

We show here a cursory exploration of the possible role of non-gravitational baryonic processes related to the release of outflows and jets from feedback sources in the simulation, which we only simulated in one of the simulated boxes of \texttt{MAKITRA}.
A full exploration of the possible role of feedback and galaxy formation is beyond the main goal of this paper, as it requires an even larger suite of simulations, including a finer scale of the many poorly constrained sub-grid model parameters connected with galaxy formation physics \citep[e.g.][and references therein for a recent review]{2025arXiv251018293M}. 

In the model variations explored here, we used the new numerical implementations discussed in \citet{Vazza2025}, after recalibrating model parameters for the coarser mass and spatial resolution of the  \texttt{MAKITRA} runs. 

 The adopted star formation recipe follows the method by  \citet[][]{2003ApJ...590L...1K} in order to reproduce the observed Kennicutt's law \citep[][]{1998ApJ...498..541K} and with free parameters calibrated to reasonably reproduce the integrated star formation history and the stellar mass function of galaxies at $z\leq 2$. 
 The feedback from star forming particles assumes a fixed fraction of energy/momentum/mass ejected per each formed star particles, $E_{SN}= \epsilon_{SF} m_* c^2$, with efficiency calibrated to $\epsilon_{SF}=10^{-8}$ as in previous work \citep[][]{va17cqg}.
 $90\%$ of the feedback energy is released in the thermal form (i.e., hot supernovae-driven winds), distributed among the 27 nearest cells around the star particle, and $10\%$ in the form of magnetic energy, assigned to magnetic dipoles by each feedback burst. 
 
The feedback from active galactic nuclei is treated at run-time by assuming that the highest density peaks in the simulation harbour a supermassive black hole, to which we attribute a realistic mass based on observed scaling relation \citep[e.g.][]{2019ApJ...884..169G}. The code computes the instantaneous mass growth rate onto each supermassive black hole by following the standard Bondi–Hoyle formalism, in which we include an ad-hoc "boost" parameter meant to compensate for the lack of resolution around the Bondi radius. Depending on the temperature of the accreted gas, the code considers either "cold gas accretion" feedback (in which most of the energy is distributed in the form of thermal energy in the neighbourhood of each simulated AGN) or "hot gas accretion" feedback (in which most of the energy is released in the form of bipolar kinetic jets).
In both cases, $10\% $ of the feedback energy is released in the form of magnetic energy, through pairs of magnetised loops wrapped around the direction of kinetic jets. This magnetic field is added at run time to any other additional magnetic field already present in the simulation. 

\begin{figure}
    \centering
              \includegraphics[width=0.45\textwidth]{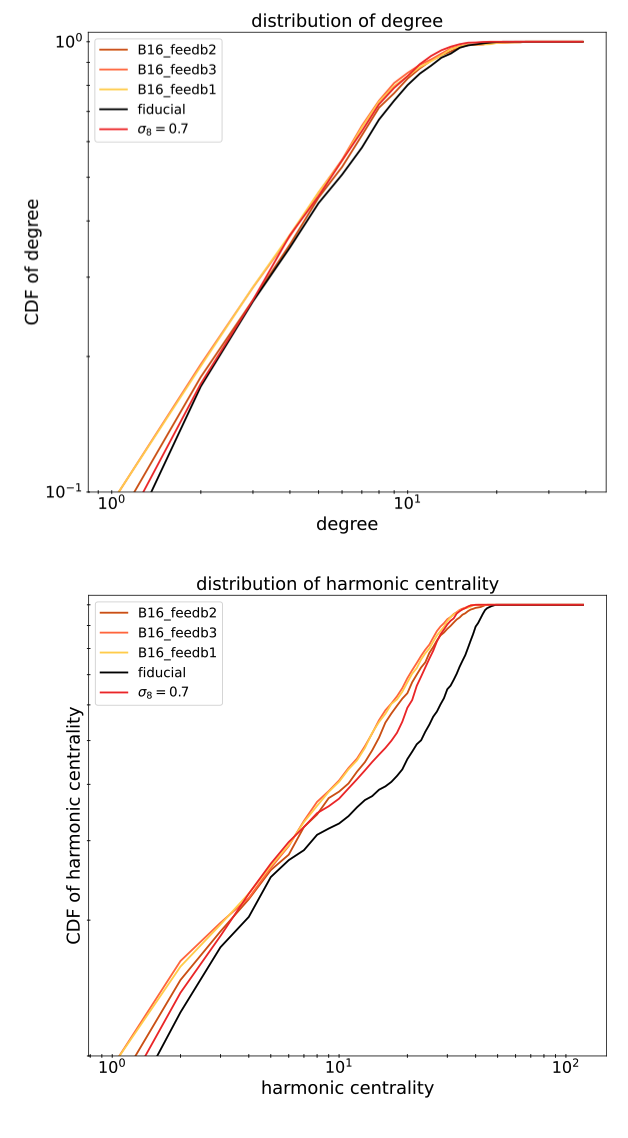}
    \caption{Cumulative distributions of degree (top panel) and harmonic centralities (bottom) for models with different values of $\sigma_8$, the fiducial $\sigma_8=0.8$ and the lower-amplitude $\sigma_8=0.7$, and for three different scenarios of baryonic feedback. All statistics are computed at $z=0.2$ using a linking length of $8~\text{Mpc}/h$. The centrality distributions correspond to a single realisation for each model (i.e., a single random seed).}
    \label{fig:distr_baryonic_feedback5}
\end{figure}

In this paper, we do not aim to systematically investigate the impact of baryonic feedback on network centrality statistics. However, to get a first impression of the amplitude of this effect, we have studied one realisation of each of three feedback models, stellar, AGN and stellar+AGN, named as \texttt{feedb\_1},\texttt{feedb\_2}, \texttt{feedb\_3} in Table~\ref{tab_cosmo}. 
Following the same procedure described in Sect.~\ref{sec:results}, we constructed the network and computed the corresponding node centrality statistics. 
For illustrative purposes, we present only the cumulative distributions of degree and harmonic centralities in Fig.~\ref{fig:distr_baryonic_feedback5}. 
Baryonic feedback effects can mimic a decrease in $\sigma_8$ of approximately 0.1 in terms of the network statistics, which is in qualitative agreement with results of \cite{Schaller:2025} for the total matter power spectra.  The introduction of the baryonic effect is likely to add to the aforementioned cosmic variance as the extra gas pressure generated by impulsive AGN feedback (and to a smaller extent, by star formation feedback) can randomly fluctuate over time for specific nodes of the network, albeit following the long-range trend of the cosmic star formation history.
For this reason,  this result should be regarded as illustrative, given that our simulation volume is again too small to fully mitigate cosmic variance effects.

\end{document}